\documentclass[runningheads]{llncs}

\usepackage{amsmath}
\usepackage{amssymb}
\usepackage{graphicx}
\usepackage{booktabs}
\usepackage[htt]{hyphenat}
\usepackage[dvipsnames]{xcolor}
\usepackage{multirow}
\usepackage{float}

\newcommand{\shahrzad}[1]{}
\newcommand{\james}[1]{}
\newcommand{\jeff}[1]{}
\newcommand{\andrew}[1]{}

\begin{document}
\title{CEQE: Contextualized Embeddings for Query Expansion}



\author{
    Shahrzad Naseri$^1$ \and
    Jeffrey Dalton$^2$ \and
    Andrew Yates$^3$ \and 
    James Allan$^1$ }
 
\authorrunning{S. Naseri et al.}

%
\institute{
    $^1$ University of Massachusetts Amherst, 
    $^2$University of Glasgow \\
    $^3$Max Planck Institute for Informatics \\
    \email{\{shnaseri, allan\}@cs.umass.edu},
    \email{jeff.dalton@glasgow.ac.uk}\\
    \email{ayates@mpi-inf.mpg.de}
}





\maketitle  

\begin{abstract}

In this work we leverage recent advances in context-sensitive language models to improve the task of query expansion. Contextualized word representation models, such as ELMo and BERT, are rapidly replacing static embedding models. We propose a new model, Contextualized Embeddings for Query Expansion (CEQE), that utilizes query-focused contextualized embedding vectors. We study the behavior of contextual representations generated for query expansion in ad-hoc document retrieval. We conduct our experiments on probabilistic retrieval models as well as in combination with neural ranking models. 
We evaluate CEQE on two standard TREC collections: Robust and Deep Learning. We find that CEQE outperforms static embedding-based expansion methods on multiple collections (by up to 18\% on Robust and 31\% on Deep Learning on average precision) and also improves over proven probabilistic pseudo-relevance feedback (PRF) models. We further find that multiple passes of expansion and reranking result in continued gains in effectiveness with CEQE-based approaches outperforming other approaches. The final model incorporating neural and CEQE-based expansion score achieves gains of up to 5\% in P@20 and 2\% in AP on Robust over the state-of-the-art transformer-based re-ranking model, Birch. 


\end{abstract}

\section{Introduction}
\label{sec:intro}
Recently there is a significant shift in text processing from high-dimensional word-based representations to ones based on continuous low-dimensional vectors. However, fundamentally both are static -- each word has a \textit{context-independent} or static representation. The fundamental challenge of polysemy remains. Recent approaches aim to address this, namely ELMo~\cite{peters-etal-2018-deep} and BERT~\cite{devlin-etal-2019-bert}, by creating \textit{context-dependent} representations that depend on the surrounding context in which they occur. The power of contextualized models comes from this ability to disambiguate and generate distinctive representations for terms with the same lexical form.  Contextualized representation models provide significant improvements across a range of diverse tasks. To our knowledge this is the first work to develop an unsupervised contextualized query expansion model based on pseudo-relevance feedback.
This represents an advancement over previous context-free expansion models based on lexical matching. Our proposed approach leverages contextual word similarity with an unsupervised expansion model. 

Contextualized representations from BERT and similar models are rapidly being adopted for retrieval and NLP, because they transfer well to new domains with limited training data. Supervised ranking models derived from them, such as CEDR~\cite{MacAvaneyYCG19} and T5~\cite{nogueira2020document}, are the top-ranked learning-to-rank methods for a wide range of retrieval and QA benchmarks.
In this work we leverage these contextualized word representations not for supervised re-ranking, but instead to improve core document matching. We address the fundamental problem that for many queries the core matching algorithms fails to identify many (or even all) relevant results in the candidate pool.  
Advancements in retrieval require more effective core matching algorithms to improve recall for neural ranking methods. No amount of reranking irrelevant results will provide relevance gains. 

We propose a new contextualized expansion method to address the task of core matching building on proven pseudo-relevance feedback (PRF) techniques from probabilistic Language Modeling and extending them to effectively leverage contextual word representations. Further, we investigate the effect of applying CEQE in combination with state-of-the-art neural re-ranking models. Our work addresses core research questions (RQ) in contextualized query expansion: 
\begin{itemize}
    \item \textbf{RQ1} How can contextualized representations be effectively leveraged to improve state-of-the-art unsupervised query expansion methods?
    \item \textbf{RQ2} How effective are neural reranking methods when performed after query expansion?
    \item \textbf{RQ3} How effective are query expansion methods after a first pass of high-precision neural re-ranking? 
\end{itemize}
We study these questions with empirical experiments on standard TREC test collections: Robust and Deep Learning 2019. The results on these test collections demonstrate that variations of CEQE significantly outperform previous static embedding models (based on GLoVe) in extrinsic retrieval effectiveness by approximately  18\% MAP on Robust04 and 31\% on TREC Deep Learning 2019 and 6-9\% for recall@1000 across all datasets.

This work makes several new contributions to methods and understanding of contextualized representations for query expansion and relevance feedback:

\begin{itemize}
\item We develop a new contextualized query expansion method, CEQE, that shifts from word-count approaches to contextualized query similarity.
\item We demonstrate through experimental evaluation that the proposed approach outperforms static embedding methods and performs at least as well as state-of-the-art word-based feedback models on multiple collections.
\item We demonstrate that neural reranking combined with CEQE results in state-of-the-art effectiveness that outperforms previous approaches.   
\end{itemize}


\section{Background and Related Work}

\textbf{Query Expansion} 
A widely used approach to improve recall uses query expansion from relevance feedback that takes a user judgment of a result's relevance and uses it to build an updated query model~\cite{rocchio-rel-feedback}. Pseudo-relevance feedback (PRF)~\cite{Lavrenko2001Relevance,lv2009comparative,zhai2001model} approaches perform this task automatically, \emph{assuming} the top documents are relevant. We build on these proven approaches based on static representations and extend them to contextualized representations.
Padaki et al.~\cite{padaki2020rethinking} investigate BERT's performance when using expanded queries and find that expansion that preserves some linguistic structure is preferrable to expanding with keywords.

\textbf{Embedding-based Expansion} Another approach for query expansion incorporates static embeddings~\cite{pennington2014glove,mikolov2013distributed} to find the relevant terms to the query, because embeddings promise to capture the semantic similarity between terms and are used in different ways to expand queries~\cite{diaz2016query,kuzi2016query,zamani2016embedding,zamani2017relevance,dalton2019local,roy2016using,naseri2018exploring}.
These word embeddings, such as Word2Vec, GloVe, and others, learn a static word embedding for each term regardless of the context. Most basic models fail to address polysemy and the contextual characteristics of terms. 
All of the previous approaches use static representations that have fundamental limitations addressed by the use of contextualized representations.

\textbf{Supervised Expansion} There is a vein of work using supervised learning to perform pseudo-relevance feedback. Cao et al. \cite{robertson-pseudo-relevance-feedback} and Imani et al.~\cite{imani2019deep} use feature-based models to try to predict what terms should be used for expansion. A common practice is to classify terms as positive, negative, or neutral and use classification methods to maximize the number of predicted positive terms. We use this labeling method to intrinsically evaluate the utility of our unsupervised approach. An end-to-end neural PRF model (NPRF) proposed by Li et al.~\cite{Li_2018} uses a combination of models to compare document summaries and compute document relevance scores for feedback and achieves limited improvement while only using bag-of-words neural models.
Later work combining BERT with a NPRF framework~\cite{zheng2020bert} illustrated the importance of an effective first-stage ranking method. A complementary vein of work ~\cite{nogueira2019document} uses generative approaches to perform \textit{document expansion} by predicting questions to add to document. In contrast, we focus on query expansion approaches. 

\textbf{Neural ranking} Contextualized Transformer-based models are now widely used for ranking tasks \cite{akkalyoncu-yilmaz-etal-2019-cross,DaiDeeperTextUnderstanding,li2020parade,MacAvaneyYCG19,nogueira2019passage,nogueira2020document,padigela2019investigating,qiao2019understanding,ZhangSXRBCT19}. 
MacAvaney et al.~\cite{MacAvaneyYCG19} propose incorporating contextualized language models into existing neural ranking architectures by considering each layer of contextualized language models as one channel and integrating the similarity matrices of each layer in the neural ranking architecture.
Recent research~\cite{gao2020complementing,Khattab_Zaharia_SIGIR2020,macavaney2020expansion,zhan2020repbert,xiong2021approximate} uses Transformer models to produce query and document representations that can be used for (relatively) efficient first-stage retrieval.
In this context, Gao et al.~\cite{gao2020complementing} find that combining a representation-based model with a lexical matching component improves effectiveness. 
In contrast, we focus on representations solely as a contextualized word representation model for the task of unsupervised query expansion.  

    
\section{Methodology}
In this section we introduce our proposed Contextualized Embedding for Query Expansion (CEQE) method that utilizes contextualized representations for the task of query expansion. The method below applies to many widely used contextualized embedding representation models, including BERT and its variants. 



\subsection{Word and WordPiece representations}
In contextualized models, to address the problem of out-of-the-vocabulary terms, subword representation such as WordPieces~\cite{Schuster2012JapaneseAK} are used. For backwards compatibility with existing word-based retrieval systems (as well as comparison with previous methods) we use words as the matching unit. We first aggregate WordPiece tokens into a contextualized vector for words. We compute the average embedding vector of word $w$ by $\overrightarrow{w} \triangleq \frac{1}{|w|}\sum_{p_i \in w} \overrightarrow{p_i}$, where $p_i$ is a WordPiece of word $w$ and $|w|$ is the number of WordPieces in the word $w$.

\subsection{Contextualized Embeddings for Query Expansion (CEQE)}
In this section we describe the core of the CEQE model. It follows in the vein of principled probabilistic language modeling approaches, such as the Relevance Model formulation of pseudo-relevance feedback \cite{Lavrenko2001Relevance}. In contrast to these approaches that are based on static lexical matching, we formulate relevance based on contextualized vector representations.
We build the contextualized feedback model based upon the core Relevance Model (RM) formulation: 
\vspace{-2pt}
\begin{align}
    p(w|\theta_R) \propto  \sum_{D\in R} p(w, Q, D)
\end{align}
\vspace{-2pt}
where $\theta_R$ and $R$ respectively denote the feedback language model and the set of pseudo-relevant documents, i.e., the top retrieved documents. In the original RM formulation, the joint probability of $p(Q,w,D)$ is broken down as follows:
\begin{align}
    \sum_{D\in R} p(w, Q, D) &= \sum_{D\in R} p(w,Q|D)p(D) \\
                             &= \sum_{D\in R} p(w|D)p(Q|D)p(D)\label{eq:independence_assumption}
\end{align}
where Equation~\ref{eq:independence_assumption} is derived from the simplifying independence assumption between the query $Q$ and term $w$. This assumption results in a static representation based on simple word counts and ignores the query explicitly. It only incorporates evidence indirectly through $P(Q|D)$. In contrast, the proposed CEQE parameterization doesn't assume term $w$ is independent of query $Q$ and explicitly incorporates the query focus based on similarity with contextualized vector representations.  More formally:
\begin{align}
    \sum_{D\in R} p(w, Q, D) = \sum_{D\in R}p(w|Q,D)p(Q|D)p(D)~\label{eq:ceqe_breakdown}
\end{align}

With a contextualized model it is no longer possible to simply count document terms -- they must be grouped, simplified, or compared against a query representation. We explicitly incorporate contextualized query similarity for each word occurrence. We now break down each of the elements in Equation~\ref{eq:ceqe_breakdown} in more detail. Following common practice we assume a uniform probability for $p(D)$.  $p(Q|D)$ is the posterior probability of the query given a document from the retrieval model. We propose several methods to calculate $p(w|Q,D)$ below.





\textbf{Centroid Representation}
In this approach we create a model of the whole query and then compare it to the contextualized representation of each word mention (occurrence), $m_w$. In the centroid representation we define $\sigma(Q)$, the aggregation of all WordPieces of the query.  Note that a representation of a query also includes special delimiter tokens. For example, in BERT this would include [CLS] and [SEP] tokens that we find carry contextual importance. We include the [CLS] token in particular because it is often used as a representation of the input with respect to the target task. For the query centroid representation we define $\sigma$ as the mean of its individual component contextual vectors: we represent query $\sigma(Q)$ by $\overrightarrow{Q} \triangleq \frac{1}{|Q|}\sum_{q_i \in Q}
\overrightarrow{q}$, where $q_i$ is a WordPiece token and $|Q|$ is the length of the query in WordPiece tokens.

We then define $p(w|Q,D)$ by comparing the similarity of individual word mentions to the query centroid representation based on a similarity function $\delta$ (e.g., cosine). If $m_w^D$ is a mention of word $w$ in a document $D$ and $M_w^D$ is the complete set of mentions of $w$: 
\vspace{-8pt}
\begin{align}
    p(w|Q,D) &\triangleq \frac{\sum_{m_w^D\in M_w^D}\delta(\overrightarrow{Q}, \overrightarrow{m_w^D})}{\sum_{m^D\in M_*^D}\delta(\overrightarrow{Q},\overrightarrow{m^D})}
\end{align}
The denominator is a normalization constant that considers all word mentions across the entire document to form a probability. This approach is novel because the contextualized vector $m_w^D$ will be different for every occurrence in $D$ because the context surrounding each mention of word $w$ varies.

\textbf{Term-based Representation} 
In this section we propose an alternative parameterization for $p(w|Q,D)$. Instead of using the centroid of the query to compute a term's similarity to the entire query, we compute the similarity for each query term separately. If $q$ is a query term and $\overrightarrow{q}$ is its corresponding contextualized embedding vector, this can be formulated as:
\vspace{-5pt}
\begin{align}
    p(w|q,D) &\triangleq \frac{\sum_{m_w^D\in M_w^D}\delta(\overrightarrow{q}, \overrightarrow{m_w^D})}{\sum_{m^D\in M^D_*}\delta(\overrightarrow{q},\overrightarrow{m^D})}
\end{align}
To select a term for expansion for the query overall we perform an extra step of pooling across the similarities of individual words.  This step combines the contextualized word vectors. Function $f$ calculate the semantic similarity of word $w$ with the whole query by combining the semantic similarity of it with each query term $q$. We define $f_{\text{max}}(w,Q,D) = \max_{q \in Q} p(w|q,D)$ and $f_{\text{prod}}(w,Q,D) = \prod_{q \in Q} p(w|q,D)$ as MaxPool and MulPool, respectively. If $Z'$ is a normalization factor that is the sum over the terms in document $D$, which is less computationally expensive than summing over all vocabulary terms, these can be defined as:
\begin{align}
    p(w|Q,D) \triangleq  \frac{f_{\text{max/prod}}(w, Q, D)}{Z'}
\end{align}


The final result of all of these methods is a relevance distribution over terms derived from the contextualized representations in top retrieved documents. The result is an updated query language model that can be used on its own or combined with other representations. 





\vspace{-7pt}
\section{Experimental Setup}
\subsection{Datasets}
We evaluate our model on two standard TREC datasets: Robust and Deep Learning.

\textbf{Robust} The corpus consists of Tipster disks 4 and 5 containing approximately 528K newswire articles. The evaluation topics are the 250 Robust topics (301-450, 601-700). We use the titles as queries.

\textbf{TREC Deep Learning}
The 2019 TREC Deep Learning (DL) Track created large labeled datasets for ad-hoc search. We perform the full document ranking task with the goal of testing new expansion methods to improve effectiveness. The evaluation has 43 test queries from Bing, and the corpus consists of 3.2 million web documents. Documents are rated on a four point graded relevance scale. The primary measure is NDCG@10.

\textbf{Evaluation Metrics} Since we focus on introducing relevant documents to a candidate pool for downstream ranking, we consider both recall-focused metrics (Recall@100, Recall@1000, MAP) as well as precision-based measures (P@10/20, NDCG@10/20). For Robust, in order to compare with previous works we report precision and NDCG at cut-off 20. We report the official primary measure for DL, NDCG@10. For significance testing, we use a paired t-test with significance at the 95\% confidence interval.
\vspace{-10pt}
\subsection{Baselines}
We study the behavior of the CEQE model in comparison with standard models from probabilistic language modeling. For the baseline retrieval we use BM25 because it is the most widely used first-pass unsupervised ranker used to generate candidate pools. We compare to two static expansion models~\cite{kuzi2016query} and a proven pseudo-relevance feedback model, the Relevance Model~\cite{Lavrenko2001Relevance}. We use the standard relevance model (RM3 variant) that performs linear interpolation of the RM expansion terms with the original query using the Query Likelihood score. 

\textbf{Static Embeddings} For static word embeddings we use GloVe~\cite{pennington2014glove} embeddings.
The pre-trained 300 dimensional Glove word embeddings are extracted from a 6 billion token collection (Wikipedia dump 2014 plus Gigawords 5). These embeddings are the most effective static embeddings for a variety of tasks, including previous work~\cite{diaz2016query} on query expansion.
We use the static embeddings with two variations. The \textit{Static-Embed model}~\cite{kuzi2016query} is a global expansion model using GloVe expansion on the target collection vocabulary. For a fair comparison with CEQE, we additionally consider a \textit{Static-Embed-PRF} variant that has its vocabulary limited to terms appearing in the PRF documents.
\vspace{-10pt}
\subsection{Intrinsic expansion judgments}
Beyond direct retrieval, we also assess term selection quality intrinsically. We directly measure the utility of individual expansion terms. Following previous work from Imani et al., we generate this term utility by performing expansion one word at a time \cite{imani2019deep}. Retrieval effectiveness assesses whether a term is good (helps retrieval), bad (hurts retrieval), or neutral (has no effect). We pool the top thousand candidate expansion terms from all candidate expansion methods. These are issued to the retrieval system with the original query (each with a default weight of 0.5, the default relevance model expansion weight). This approach follows standard relevance model interpolation practice and removes the dependence on the original query length (instead of simply appending a word). We measure improvement based on recall@1000 with a threshold of 0.001.
For Robust this results in approximately 500k candidate terms. For the intrinsic evaluation only queries with at least one positive expansion term are used. This is 181 queries for Robust with 10,068 positive terms. 
\subsection{System Details}
All collections are indexed with the Galago\footnote{\url{http://www.lemurproject.org/galago.php}} open-source retrieval system for research. The query models and feedback expansion models are all implemented using the Galago query language. We perform stopword removal and stemming using Galago's stopword list and Krovetz stemmer, respectively. 



\textbf{Contextualized Embedding Model} We use BERT because it is the most widely used contextual representation model. We use the pre-trained BERT (BERT-Base, Uncased) model with maximum sequence length of 128 for calculating the contextualized embedding vectors. Since the documents in Robust are longer than 128 tokens we split the documents into chunks with maximum size of 128 tokens. For the primary CEQE results in this section we use a single layer of the contextualized representation, the second to last layer (11) of BERT. This layer was shown to be the most effective single layer on NER \cite{devlin-etal-2019-bert} and it was shown that later layers (before the last) were the most effective word representations for multiple language tasks \cite{Peters2019ToTO} that use contextual embeddings as features. Initial preliminary experiments confirmed this finding.

\textbf{Neural ranking models}
For our neural models we adopt CEDR~\cite{MacAvaneyYCG19}. In particular, to align with the use of the contextualized models we use the BERT variant.  For Robust, we use the CEDR-KNRM model trained by the authors~\cite{MacAvaneyYCG19}. Throughout the paper we refer to the CEDR-KNRM as CEDR. For DL we use a CEDR variant trained on a random sample of 1000 MS MARCO train queries with early stopping to terminate when there is no validation improvement for 20 iterations. 

\textbf{Parameter settings}
The unsupervised retrieval and feedback hyperparameters are tuned using grid search. The $b$ and $k1$ are tuned for BM25 as well as $mu$ for the QL model in the RM3 score. For all PRF query expansion methods we tune the number of documents ($\{5, 10,..., 100\}$ by 5), terms ($\{10, 20,..., 100\}$ by 10) , and interpolation coefficient ($\{0.1, 0.2, ..., 0.9\}$ by 0.05). For Robust, we use  five-fold cross-validation with the splits introduced by Huston and Croft~\cite{huston2014parameters}. 
For DL the original 2019 track only used MS MARCO for training. We set hyper-parameters using five cross-validation with random splits on the topics. 

\section{Experimental Results}
First, in Section~\ref{sec:intrinsic-exp} we study how to incorporate contextualized embeddings for the task of unsupervised query expansion (RQ1). Then, in Section~\ref{sec:feedback-neural-reranker} we explore the effect of CEQE variants in combination with neural ranking methods (RQ2). Finally, in Section~\ref{sec:neural-reranking-expansion} section we study how a reranked neural result can be used as a basis for further expansion and reranking (RQ3).

\vspace{-10pt}
\subsection{Unsupervised Expansion Comparison}
\label{sec:unsupervised-exp}
We first evaluate our expansion model on retrieval effectiveness in extrinsic evaluation. We study this setup because these are the most widely used algorithms for first pass retrieval. In this pass it is critical to focus on recall at a cutoff, particularly with a low cutoff due to the computational requirements of second pass reranking (e.g., the top 100 documents as in \cite{MacAvaneyYCG19} and the Deep Learning Track \cite{2019trecdl}). We present the results of the methods as well as baselines for Robust04 in Table~\ref{tab:robust-unsupervised} and 2019 Deep Learning Track in Table~\ref{tab:dl-unsupervised}. 

\textbf{Robust (Table~\ref{tab:robust-unsupervised})}
The results on Robust show that all expansion methods outperform the baseline BM25 retrieval method across all measures. The static embedding models outperform BM25, but do not perform as well as the Relevance Model (RM3). The effectiveness of the Static-Embed-PRF method that only uses terms in the PRF documents' vocabulary is more effective across all measures over the Static-Embed approach with a global vocabulary. We hypothesize that this may be due to the fact that the query results provide a topically focused vocabulary and filters out generally similar noise. RM3 significantly outperforms the Static-Embed method for MAP, but not other measures. 
To give an indicator of the  BM25 + RM3 parameters, the average parameter settings across the folds is: 22 feedback docs, 71 expansion terms, and interpolation weight of 0.3.
We observe that the contextualized expansion methods outperform the static embedding models. 
The results show the best method is CEQE-MaxPool. The Centroid method is slightly lower than MaxPool, and both outperform multiplicative pooling. The CEQE-MaxPool result outperforms the BM25+RM3 across all measures and in Recall@1000 is significant over both static embedding methods and BM25+RM3, which demonstrates the utility of context-dependent embeddings. 

\begin{table*}[t]
    \centering
    \caption{Ranking effectiveness on the Robust collection. The superscript $\dagger$ and $\ddagger$ denotes statistical significance over BM25 + RM3 and Static-Embed-PRF, respectively.}
    \label{tab:robust-unsupervised}
    \scalebox{0.95}{
    \begin{tabular}{l ccccccc}
        \toprule
        \textbf{Model} & P@20 & nDCG@20 & mAP@1000 & Recall@100 & Recall@1000\\
        \midrule
        BM25                & 0.3657 & 0.4193 & 0.2574 & 0.4165 & 0.6933 \\
        BM25 + RM3          & 0.3998 & 0.4517  & 0.3069 & 0.4610$^\ddagger$ & 0.7588$^\ddagger$ \\
        Static-Embed        & 0.3675 & 0.4285 & 0.2615 & 0.4217 & 0.7125\\ 
        Static-Embed-PRF    & 0.3781 & 0.4400 & 0.2703 & 0.4324 & 0.7231\\
        CEQE-Centroid        &  0.3922 & 0.4462 &  0.3019$^\ddagger$ &  0.4593$^\ddagger$ &  0.7653$^{\dagger\ddagger}$ \\
         CEQE-MulPool        & 0.3847 & 0.4360 & 0.2845$^\ddagger$ & 0.4517$^\ddagger$ & 0.7435$^\ddagger$ \\
        CEQE-MaxPool        & \textbf{0.4040}$^\ddagger$ & \textbf{0.4587} & \textbf{0.3086}$^\ddagger$ &\textbf{ 0.4651}$^\ddagger$ &  \textbf{0.7689}$^{\dagger\ddagger}$ \\
        \midrule
        CEQE-MaxPool(fine-tuned)  & 0.3986$^\ddagger$ & 0.4528 & 0.3071$^\ddagger$ & 0.4647$^\ddagger$ & 0.7626$^\ddagger$& \\
        \bottomrule
    \end{tabular}
    }
    \vspace{-10pt}
\end{table*}

The last line of the table shows the result of using MaxPool with `fine-tuned' contextual embeddings from a BERT model trained for ranking on Robust. The results show small and insignificant differences across all measures. It is almost identical to vanilla embedding effectiveness after being combined with RM3. This indicates that, when used for CEQE-based expansion, pre-trained models are comparable in effectiveness to ones fine-tuned for ranking.
\shahrzad{is it True?} To our knowledge these are the best unsupervised query expansion results for Robust that do not use external collections.

\textbf{Deep Learning 19 (Table~\ref{tab:dl-unsupervised})}
We report the official evaluation measures for the TREC 2019 Deep Learning Track \cite{2019trecdl} as well as Recall@1000. For NDCG@10, the baseline BM25 retrieval is more effective than all expansion methods. To give an indicator of the BM25 + RM3 parameters, the average parameter settings across the folds is: 15 feedback docs, 85 expansion terms, and interpolation weight of 0.4. Similar to Robust, we observe that a tuned RM3 outperforms the static embedding methods across all measures. CEQE-MulPool and CEQE-MaxPool also outperform the static embedding model across all measures. The best performing expansion method is CEQE-MaxPool, outperforming RM3. 
We note that given the small sample size (43 topics), none of the unsupervised methods show statistically significant differences between them. As shown later, that requires performing expansion on top of neural rankings.

Although our experimental setup is based on cross-fold validation (rather than tuning on MARCO), we include the reported values from the Deep Learning track overview \cite{2019trecdl} for reference. Importantly, we observe that the CEQE-MaxPool outperforms all submitted TREC systems on recall@1000 and is in the top five for recall@100. It's noteworthy that the unsupervised CEQE-MaxPool `traditional' model is only slightly lower than the median for P@10 and NDCG@10 with runs that include many state-of-the-art neural models.

\begin{table}[t]
    \centering
    \caption{Ranking effectiveness of CEQE on unsupervised baseline retrieval for Deep Learning 2019 Track for the task of full document ranking. The superscript $\dagger$ and $\ddagger$ denotes statistical significance over BM25 + RM3 and Static-Embed, respectively. }
    \label{tab:dl-unsupervised}
    \scalebox{0.95}{
    \begin{tabular}{l ccccccc}
        \toprule
        \textbf{Model} & P@10 & nDCG@10 & mAP@1000 & Recall@100 & Recall@1000 \\
        \midrule
        BM25                & 0.6535 & \textbf{0.5730} & 0.3513 & 0.4053 & 0.6950\\
        BM25 + RM3          & 0.6256 & 0.5343 & 0.3975$^\ddagger$ & 0.4434$^\ddagger$ & 0.7750$^\ddagger$\\
         Static-Embed        & 0.6186 & 0.5427 & 0.3373 & 0.3973 & 0.7179 & \\ 
        Static-Embed-PRF    & 0.5605 & 0.4925 & 0.3166 & 0.3715 & 0.6737  \\
      CEQE-Centroid        & 0.5580 & 0.5580 & 0.4144$^\ddagger$ & 0.4464$^\ddagger$ & 0.7804$^\ddagger$  \\
      CEQE-MulPool        & 0.6442 & 0.5563 & 0.3724$^\ddagger$ & 0.4295$^\ddagger$ & 0.7560$^\ddagger$  \\        
        CEQE-MaxPool        & \textbf{0.6581} & 0.5614 & 0.4161$^{\dagger\ddagger}$ & 0.4506$^\ddagger$ & 0.7832$^\ddagger$  \\
            \midrule
             \midrule
         TREC 2019 Median & 0.6597 & 0.5834 & 0.2984 & 0.3748 & 0.5484\\
         TREC 2019 Best & 0.8093 & 0.7260 & 0.4280 & 0.4670 & 0.7553\\
        \bottomrule
    \end{tabular}
    }
    \vspace{-5pt}
\end{table}

\vspace{-10pt}
\subsubsection{Intrinsic Evaluation}
\label{sec:intrinsic-exp}
In this section we examine the effectiveness of the expansion approaches to rank positive expansion terms that improve Mean Average Precision (at 1000) when added to the query. This experiment evaluates a method's ability to identify good expansion terms in isolation. The results are shown in Table~\ref{tab:intrinsic-eval} for the key expansion models to compare for Robust collection. Because a fixed top-k expansion terms are usually selected for expansion we evaluate the intrinsic evaluation with set-based precision numbers at common thresholds for the number of expansion terms. The results show that a well-tuned Relevance Model outperforms query expansion models based on static embeddings. In contrast, we find that our proposed contextualized embedding model, CEQE, provide improvements in early ranks for P@10 and P@20. All the CEQE models significantly improve over static embedding models across all metrics. And further, we find that CEQE-MaxPool significantly outperforms the Relevance Model expansion effectiveness for P@10 and P@20. It is insignificantly different from the Relevance Model at rank 100. This indicates that strength of CEQE is selecting a higher number of ``good'' terms earlier, allowing improved effectiveness with fewer expansion terms.

\begin{table}[t]
    \centering
    \caption{Intrinsic ranking evaluation of expansion terms on Robust. Significance over Relevance Model is indicated by $\dagger$ and Static-Embed-PRF by $\ddagger$.}
    \label{tab:intrinsic-eval}
  \scalebox{0.90}{%
    \begin{tabular}{l ccc}
        \toprule
        \textbf{Model} & P@10 & P@20 & P@100\\
        \midrule
        Relevance Model                      & 0.1693$^{\ddagger}$ & 0.1419$^\ddagger$ & \textbf{0.0871}$^{\ddagger}$\\
        Static-Embed             & 0.1008 & 0.0780 & 0.0511\\
        Static-Embed-PRF         & 0.1357 & 0.1083 & 0.0655 \\
        CEQE-MulPool & 0.1349 & 0.1174 & 0.0737\\
        CEQE-Centroid            & 0.1751$^{\ddagger}$ & 0.1481$^{\ddagger}$ & 0.0826$^{\ddagger}$\\
        CEQE-MaxPool             &\textbf{0.1830}$^{\dagger\ddagger}$ & \textbf{0.1500}$^{\dagger\ddagger}$ & 0.0841$^{\ddagger}$\\
        \midrule
    \end{tabular}
    }
    \vspace{-20pt}
\end{table}

We explore the intrinsic results in more depth with an example for one topic on Robust in Table~\ref{tab:example-queries}. 
The first column has the terms (unstemmed) with the greatest improvement for the query. The ranking of expansion terms for the Relevance Model and CEQE-MaxPool are shown for comparison. We observe that CEQE model identifies all of the terms from RM as well as three additional relevant terms. More generally, we see that the CEQE terms appear to have a stronger semantic relationship with the query terms. The RM terms appear most loosely related and have additional noise terms, including single digit numbers. This is because RM focuses on terms that co-occur across multiple PRF documents, but it does not explicitly model the relationship to the query. In contrast our proposed model explicitly focuses on the query. As a result, the CEQE model produces fewer terms that co-occur by chance.

\begin{table}[t]
    \centering
    \caption{Example top expansion terms for Topic 685, [oscar winner selection]. This includes a sample of the most important intrinsic positive labels, Relevance Model terms, and CEQE Expansion terms. Terms with positive intrinsic labels are highlighted. }
    \label{tab:example-queries}
    \scalebox{0.9}{
    \begin{tabular}{rl}
    \toprule
\textbf{Positive terms:} & academy, academys, nominations, nomination, critics, members, \\
& branch, ignored, true, films, film, directors, director, filmmaker \\
\midrule
\textbf{RM:} & best, \textbf{film}, picture, million, \textbf{academy}, years, \textbf{award}, home, \\
& edition, \textbf{films}, man, four, 1, 5 \\
\midrule
\textbf{CEQE-Maxpool:} & \textbf{film}, \textbf{academy}, picture, winners, \textbf{award}, \textbf{films}, million, oscars, \\
& box, presented, \textbf{awards}, \textbf{director}, years, \textbf{nominations} \\
        \bottomrule
    \end{tabular}
    }
\end{table}

\vspace{-10pt}
\subsection{PRF effect on Neural Reranking}
\label{sec:feedback-neural-reranker}
\begin{table*}[t]
    \centering
    \caption{Ranking effectiveness of neural ranking on top of query expansion methods for Robust. The superscript $\dagger$ and $\ddagger$ indicates significance over BM25 + CEDR and (BM25 + RM3) + CEDR with re-ranking top 1000, respectively. }
    \resizebox{\columnwidth}{!}{%
    \begin{tabular}{l ccccccc}
        \toprule
        \textbf{Model} & P@20 & nDCG@20 & mAP@1000 & Recall@100 & Recall@1000\\
        \midrule
         BM25 + RM3          & 0.3998 & 0.4517 & 0.3069 & 0.4610 & 0.7588 \\
        BM25 + CEDR \cite{MacAvaneyYCG19}     & 0.4713 & 0.5458 & 0.3312 & 0.4983 & 0.6933 \\
        (BM25 + RM3) + CEDR   & 0.4719 & 0.5435 & 0.3500$^\dagger$ & 0.5192$^\dagger$ & 0.7570$^\dagger$\\ 
        (BM25 + CEQE-MaxPool) + CEDR   & \textbf{0.4735} & \textbf{0.5462} & \textbf{0.3532}$^\dagger$ & \textbf{0.5258}$^{\dagger\ddagger}$ & \textbf{0.7719}$^{\dagger\ddagger}$\\ 
      
        \bottomrule
    \end{tabular}
    }
    \label{tab:dl-supervised-exp}
    \vspace{-13pt}
\end{table*}
We now study how PRF methods impact the effectiveness of neural reranking models (RQ2). It is important to have effective expansion in the first pass to retrieve sufficient numbers of documents to rerank. The results of our experiments on Robust are shown in Table~\ref{tab:dl-supervised-exp}. 
Applying neural reranked models baselines designed for document ranking, CEDR~\cite{MacAvaneyYCG19}, on expanded query runs results in significant gains to average precision, recall@100, and recall@1000 for both RM3 and CEQE. Replacing RM3 with CEQE for expansion results in significant improvement over Recall@100 and Recall@1000. The PRF parameters are 20 documents, 90 terms, and interpolation weight of 0.3. 
\vspace{-7pt}
\subsection{Expansion after Reranking}

In this section we study how a reranked neural result can be used as a basis for further expansion and reranking (RQ3). This is a critical step because there must be a sufficient number of relevant documents in the top ranks for PRF to be effective. We evaluate multi-round supervised reranking based on expansion runs for Robust in Table~\ref{tab:robust-expansion-rerank}. The top of the table shows results from the leading neural ranking and PRF approaches, including Neural PRF~\cite{Li_2018}, CEDR, and Birch~\cite{yilmaz2019applying}. 
The results in this section all perform re-ranking on 1000 results from the baseline. We experimented with reranking 100 results and found it consistently performed worse. 
The baseline model run is BM25+CEDR followed by RM3 expansion with CEDR reranking, which we denote as \textit{(BM25 + CEDR) + RM3 + CEDR}. The results show it outperforms Birch in NDCG@20 and P@20, as well as its own previous result for P@20 on just BM25. Replacing RM3 with CEQE for the expansion consistently outperforms the previous best CEDR results across all measures and significantly over Recall@1000. 
The runs compare performing RM3 and CEQE-MaxPool on the CEDR baseline (which reranks an initial BM25 first run).  The second pass results are then reranked again using CEDR. The result is further improve over previous approaches. The same trend continues, with the CEQE-MaxPool outperforming the reranked RM3 run. 

A common approach when applying BERT-based neural ranking is to perform learning-to-rank to combine the BERT and retrieval score.  A simple proven approach is the linear interpolation of the underlying retrieval score with neural ranking model \cite{yilmaz2019applying,yang2019end}. We apply this to the two best runs, learning the interpolation using the previously described cross-validation setup. The results demonstrate that linear interpolation with these expansion runs continues to show gains. The interpolation with CEQE-MaxPool is the best performing, and compared with the previous Birch shows over 5\% relative gain P@20 and nDCG@20 as well as improving MAP.
These results show that multiple rounds of expansion and reranking can continue to result in significant improvements.

\label{sec:neural-reranking-expansion}
\begin{table*}[t]
    \centering
    \caption{Ranking effectiveness of multi-round neural re-ranking and expansion for Robust. The superscript $\dagger$ and $\ddagger$ indicates significance over BM25 + CEDR and (BM25 + CEDR) + RM3 baselines, respectively.}
    \label{tab:robust-expansion-rerank}
    \resizebox{\columnwidth}{!}{%
    \begin{tabular}{l ccccc}
        \toprule
        \textbf{Model} & P@20 & nDCG@20 & mAP@1000 & Recall@100 & Recall@1000\\
        \midrule
        Neural PRF-DRMM \cite{Li_2018}      & 0.4064 & 0.4576 & 0.2904 & - & -\\ 
        BM25 + CEDR     \cite{MacAvaneyYCG19}     & 0.4713 & 0.5458 & 0.3312 & 0.4983 & 0.6933\\
        Birch \cite{yilmaz2019applying}  &0.4657  & 0.5325 & 0.3697 & - & -\\
        \midrule
        \midrule
        (BM25 + CEDR) + RM3    & 0.4458 & 0.5211 &  0.3321 & 0.4881 &0.7751$^\dagger$\\
        (BM25 + CEDR) + RM3 + CEDR   & 0.4783 & 0.5499 &  0.3574$^\dagger$ & 0.5291$^\dagger$ & 0.7751$^\dagger$\\
        (BM25 + CEDR) + RM3 + CEDR Interp  & 0.4837$^\dagger$ & 0.5565 &   0.3739$^\dagger$ & 0.5440$^\dagger$& 0.7751$^\dagger$\\
        \hline
         (BM25 + CEDR) + CEQE-MaxPool    & 0.4504 & 0.5250 &  0.3366 & 0.4931 & 0.7874$^{\dagger\ddagger}$\\
         
        (BM25 + CEDR) + CEQE-MaxPool + CEDR   & 0.4799 & 0.5516  & 0.3601$^\dagger$ & 0.5332$^\dagger$ & 0.7874$^{\dagger\ddagger}$\\
        (BM25 + CEDR) + CEQE-MaxPool + CEDR Interp   & \textbf{0.4904}$^\dagger$ & \textbf{0.5621}$^\dagger$  & \textbf{0.3773}$^\dagger$ & \textbf{0.5486}$^\dagger$ & \textbf{0.7874}$^{\dagger\ddagger}$\\
        \bottomrule
    \end{tabular}
    }
    \vspace{-10pt}
\end{table*}
\vspace{-10pt}
\section{Conclusion}
\vspace{-2pt}
We introduce a new method, CEQE, for query expansion that extends relevance feedback approaches to recent advances in contextualized language models. CEQE address fundamental challenges using context-dependent term representations for unsupervised pseudo-relevance feedback. We study its empirical effectiveness on multiple standard test collections and the results demonstrate that they are superior to previous static embedding approaches. 

\vspace{-5pt}
\section*{Acknowledgement}
This work was supported in part by the Center for Intelligent Information Retrieval and in part by NSF grant \#IIS-1617408. Any opinions, findings and conclusions or recommendations expressed in this material are those of the authors and do not necessarily reflect those of the sponsor.

\newpage
\bibliographystyle{splncs04}
\bibliography{bibliography}

\begin{thebibliography}{10}
\providecommand{\url}[1]{\texttt{#1}}
\providecommand{\urlprefix}{URL }
\providecommand{\doi}[1]{https://doi.org/#1}

\bibitem{akkalyoncu-yilmaz-etal-2019-cross}
Akkalyoncu~Yilmaz, Z., Yang, W., Zhang, H., Lin, J.: Cross-domain modeling of
  sentence-level evidence for document retrieval. In: Proceedings of the 2019
  Conference on Empirical Methods in Natural Language Processing and the 9th
  International Joint Conference on Natural Language Processing (EMNLP-IJCNLP).
  Association for Computational Linguistics, Hong Kong, China (Nov 2019)

\bibitem{robertson-pseudo-relevance-feedback}
Cao, G., Nie, J.Y., Gao, J., Robertson, S.: {Selecting good expansion terms for
  pseudo-relevance feedback}. In: Proceedings of the 31st annual international
  ACM SIGIR conference on Research and development in information retrieval.
  SIGIR '08, ACM, New York, NY, USA (2008)

\bibitem{2019trecdl}
Craswell, N., Mitra, B., Yilmaz, E., Campos, D.: Overview of the trec 2019 deep
  learning track. In: Proceedings of The Twenty-Eight Text REtrieval
  Conference, {TREC} 2019, Gaithersburg, Maryland, USA, November 13-15, 2019
  (2019)

\bibitem{DaiDeeperTextUnderstanding}
Dai, Z., Callan, J.: Deeper text understanding for ir with contextual neural
  language modeling. In: Proceedings of the 42nd International ACM SIGIR
  Conference on Research and Development in Information Retrieval. SIGIR’19,
  Association for Computing Machinery, New York, NY, USA (2019)

\bibitem{dalton2019local}
Dalton, J., Naseri, S., Dietz, L., Allan, J.: Local and global query expansion
  for hierarchical complex topics. In: European Conference on Information
  Retrieval. pp. 290--303. Springer (2019)

\bibitem{devlin-etal-2019-bert}
Devlin, J., Chang, M.W., Lee, K., Toutanova, K.: {BERT}: Pre-training of deep
  bidirectional transformers for language understanding. In: Proceedings of the
  2019 Conference of the North {A}merican Chapter of the Association for
  Computational Linguistics: Human Language Technologies, Volume 1 (Long and
  Short Papers). Association for Computational Linguistics, Minneapolis,
  Minnesota (Jun 2019)

\bibitem{diaz2016query}
Diaz, F., Mitra, B., Craswell, N.: Query expansion with locally-trained word
  embeddings. In: Proceedings of the 54th Annual Meeting of the Association for
  Computational Linguistics (Volume 1: Long Papers) (2016)

\bibitem{gao2020complementing}
Gao, L., Dai, Z., Fan, Z., Callan, J.: Complementing lexical retrieval with
  semantic residual embedding. arXiv preprint arXiv:2004.13969  (2020)

\bibitem{huston2014parameters}
Huston, S., Croft, W.B.: Parameters learned in the comparison of retrieval
  models using term dependencies. Ir, University of Massachusetts  (2014)

\bibitem{imani2019deep}
Imani, A., Vakili, A., Montazer, A., Shakery, A.: Deep neural networks for
  query expansion using word embeddings. In: European Conference on Information
  Retrieval. pp. 203--210. Springer (2019)

\bibitem{Khattab_Zaharia_SIGIR2020}
Khattab, O., Zaharia, M.: {ColBERT}: Efficient and effective passage search via
  contextualized late interaction over {BERT}. In: Proceedings of the 43rd
  Annual International ACM SIGIR Conference on Research and Development in
  Information Retrieval (SIGIR 2020) (2020)

\bibitem{kuzi2016query}
Kuzi, S., Shtok, A., Kurland, O.: Query expansion using word embeddings. In:
  Proceedings of the 25th ACM international on conference on information and
  knowledge management. ACM (2016)

\bibitem{Lavrenko2001Relevance}
Lavrenko, V., Croft, W.B.: {Relevance based language models}. In: Proceedings
  of the 24th annual international ACM SIGIR conference on Research and
  development in information retrieval. SIGIR '01, ACM, New York, NY, USA
  (2001)

\bibitem{Li_2018}
Li, C., Sun, Y., He, B., Wang, L., Hui, K., Yates, A., Sun, L., Xu, J.: Nprf: A
  neural pseudo relevance feedback framework for ad-hoc information retrieval.
  Proceedings of the 2018 Conference on Empirical Methods in Natural Language
  Processing  (2018)

\bibitem{li2020parade}
Li, C., Yates, A., MacAvaney, S., He, B., Sun, Y.: Parade: Passage
  representation aggregation for document reranking. arXiv preprint
  arXiv:2008.09093  (2020)

\bibitem{lv2009comparative}
Lv, Y., Zhai, C.: A comparative study of methods for estimating query language
  models with pseudo feedback. In: Proceedings of the 18th ACM conference on
  Information and knowledge management. ACM (2009)

\bibitem{macavaney2020expansion}
MacAvaney, S., Nardini, F.M., Perego, R., Tonellotto, N., Goharian, N.,
  Frieder, O.: Expansion via prediction of importance with contextualization.
  arXiv preprint arXiv:2004.14245  (2020)

\bibitem{MacAvaneyYCG19}
MacAvaney, S., Yates, A., Cohan, A., Goharian, N.: {CEDR:} contextualized
  embeddings for document ranking. In: Proceedings of the 42nd International
  {ACM} {SIGIR} Conference on Research and Development in Information
  Retrieval, {SIGIR} 2019, Paris, France, July 21-25, 2019. (2019)

\bibitem{mikolov2013distributed}
Mikolov, T., Sutskever, I., Chen, K., Corrado, G.S., Dean, J.: Distributed
  representations of words and phrases and their compositionality. In: Advances
  in neural information processing systems (2013)

\bibitem{naseri2018exploring}
Naseri~S., Foley~J., A.J., B., O.: Exploring summary-expanded entity embeddings
  for entity retrieval. In: CEUR Workshop Proceedings (2018)

\bibitem{nogueira2019passage}
Nogueira, R., Cho, K.: Passage re-ranking with bert. arXiv preprint
  arXiv:1901.04085  (2019)

\bibitem{nogueira2020document}
Nogueira, R., Jiang, Z., Lin, J.: Document ranking with a pretrained
  sequence-to-sequence model. arXiv preprint arXiv:2003.06713  (2020)

\bibitem{nogueira2019document}
Nogueira, R., Yang, W., Lin, J., Cho, K.: Document expansion by query
  prediction. arXiv preprint arXiv:1904.08375  (2019)

\bibitem{padaki2020rethinking}
Padaki, R., Dai, Z., Callan, J.: Rethinking query expansion for bert reranking.
  In: European Conference on Information Retrieval. Springer (2020)

\bibitem{padigela2019investigating}
Padigela, H., Zamani, H., Croft, W.B.: Investigating the successes and failures
  of bert for passage re-ranking. arXiv preprint arXiv:1905.01758  (2019)

\bibitem{pennington2014glove}
Pennington, J., Socher, R., Manning, C.: Glove: Global vectors for word
  representation. In: Proceedings of the 2014 conference on empirical methods
  in natural language processing (EMNLP) (2014)

\bibitem{peters-etal-2018-deep}
Peters, M., Neumann, M., Iyyer, M., Gardner, M., Clark, C., Lee, K.,
  Zettlemoyer, L.: Deep contextualized word representations. In: Proceedings of
  the 2018 Conference of the North {A}merican Chapter of the Association for
  Computational Linguistics: Human Language Technologies, Volume 1 (Long
  Papers). Association for Computational Linguistics, New Orleans, Louisiana
  (Jun 2018)

\bibitem{Peters2019ToTO}
Peters, M.E., Ruder, S., Smith, N.A.: To tune or not to tune? adapting
  pretrained representations to diverse tasks. In: RepL4NLP@ACL (2019)

\bibitem{qiao2019understanding}
Qiao, Y., Xiong, C., Liu, Z., Liu, Z.: Understanding the behaviors of bert in
  ranking. arXiv preprint arXiv:1904.07531  (2019)

\bibitem{rocchio-rel-feedback}
Rocchio, J.J.: {Relevance feedback in information retrieval}. In: Salton, G.
  (ed.) The SMART Retrieval System: Experiments in Automatic Document
  Processing, chap.~14, pp. 313--323. Prentice-Hall Series in Automatic
  Computation, Prentice-Hall, Englewood Cliffs NJ (1971)

\bibitem{roy2016using}
Roy, D., Paul, D., Mitra, M., Garain, U.: Using word embeddings for automatic
  query expansion  (July 2016)

\bibitem{Schuster2012JapaneseAK}
Schuster, M., Nakajima, K.: Japanese and korean voice search. 2012 IEEE
  International Conference on Acoustics, Speech and Signal Processing (ICASSP)
  (2012)

\bibitem{xiong2021approximate}
Xiong, L., Xiong, C., Li, Y., Tang, K.F., Liu, J., Bennett, P.N., Ahmed, J.,
  Overwikj, A.: Approximate nearest neighbor negative contrastive learning for
  dense text retrieval. In: International Conference on Learning
  Representations (2021)

\bibitem{yang2019end}
Yang, W., Xie, Y., Lin, A., Li, X., Tan, L., Xiong, K., Li, M., Lin, J.:
  End-to-end open-domain question answering with bertserini. In: Proceedings of
  the 2019 Conference of the North American Chapter of the Association for
  Computational Linguistics (Demonstrations) (2019)

\bibitem{yilmaz2019applying}
Yilmaz, Z.A., Wang, S., Yang, W., Zhang, H., Lin, J.: Applying bert to document
  retrieval with birch. In: Proceedings of the 2019 Conference on Empirical
  Methods in Natural Language Processing and the 9th International Joint
  Conference on Natural Language Processing (EMNLP-IJCNLP): System
  Demonstrations. pp. 19--24 (2019)

\bibitem{zamani2016embedding}
Zamani, H., Croft, W.B.: Embedding-based query language models. In: Proceedings
  of the 2016 ACM international conference on the theory of information
  retrieval. ACM (2016)

\bibitem{zamani2017relevance}
Zamani, H., Croft, W.B.: Relevance-based word embedding. In: Proceedings of the
  40th International ACM SIGIR Conference on Research and Development in
  Information Retrieval. pp. 505--514. ACM (2017)

\bibitem{zhai2001model}
Zhai, C., Lafferty, J.: Model-based feedback in the language modeling approach
  to information retrieval. In: Proceedings of the tenth international
  conference on Information and knowledge management. CIKM '01, ACM, ACM, New
  York, NY, USA (2001)

\bibitem{zhan2020repbert}
Zhan, J., Mao, J., Liu, Y., Zhang, M., Ma, S.: Repbert: Contextualized text
  embeddings for first-stage retrieval. arXiv preprint arXiv:2006.15498  (2020)

\bibitem{ZhangSXRBCT19}
Zhang, H., Song, X., Xiong, C., Rosset, C., Bennett, P.N., Craswell, N.,
  Tiwary, S.: Generic intent representation in web search. In: Proceedings of
  the 42nd International {ACM} {SIGIR} Conference on Research and Development
  in Information Retrieval, {SIGIR} 2019, Paris, France, July 21-25, 2019.
  (2019)

\bibitem{zheng2020bert}
Zheng, Z., Hui, K., He, B., Han, X., Sun, L., Yates, A.: Bert-qe:
  Contextualized query expansion for document re-ranking. In: Proceedings of
  the 2020 Conference on Empirical Methods in Natural Language Processing:
  Findings. pp. 4718--4728 (2020)

\end{thebibliography}

\end{document}